# Interagir avec un contenu opératique : le projet d'opéra virtuel interactif *Virtualis*

**Alain Bonardi**
Université Paris IV
alain.bonardi@wanadoo.fr

**Francis Rousseaux**
Université de Reims
francis.rousseaux@univ-reims.fr

**Résumé**

Dans cet article, nous présentons le projet d'opéra interactif sur CD-ROM *Virtualis*. Ce projet comporte une dimension scientifique aussi bien qu'artistique. Il nous a donné l'occasion de concevoir une modélisation de la représentation d'opéra en utilisant des formalismes des sciences de l'organisation. De plus, notre réflexion sur l'interaction entre un utilisateur et des contenus opératiques nous a conduits à utiliser des modèles de relations entre entités fondés sur des forces physiques, dont l'utilisateur est en quelque sorte absent. Nous détaillons quelques aspects de cet environnement de lecture mais aussi d'écriture sur des contenus artistiques complexes, entre texte, musique et graphiques.

## 1. Principes de *Virtualis*

### 1.1. Introduction

*Virtualis* est un opéra interactif sur support informatique (CD-ROM) que nous développons à la fois comme projet artistique de création, initié par Alain Bonardi, et comme projet de recherche dans le domaine de l'interaction homme-machine, impulsé par Francis Rousseaux. Il offre à un spectateur la possibilité de jouer[1] non pas *à l'opéra* – il ne s'agit pas d'une reconstitution en réalité virtuelle d'une œuvre du passé ou d'une salle lyrique existante –, mais bien *avec l'opéra*, c'est-à-dire d'interagir avec des contenus lyriques. Dans le cas d'un opéra numérique tel que *Virtualis*, ces derniers recouvrent des textes (à lire ou enregistrés – lus ou chantés –), des fragments musicaux et sonores, ainsi que des éléments graphiques.

---

[1] Pour l'instant, *Virtualis* n'offre pas de possibilités de « jeu d'opéra » en réseau.

Si virtuel il y a, est visé celui de la dyade actuel/virtuel plutôt que celui de l'expression « réalité virtuelle » [Bonardi & Rousseaux 1999]. Nous recherchons pour notre cadre informatique multimédia des procédures qui ne se satisfairaient pas de simples moyens combinatoires d'aiguillage d'une possibilité à l'autre, mais tenteraient de s'appuyer sur un modèle de résolution de problème, autrement dit d'actualisation. Toutefois, dans le cadre de *Virtualis*, les problèmes à résoudre seront bien spécifiques, puisqu'ils concerneront la mise en forme de contenus musicaux et narratifs. Par exemple, l'un des problèmes posés est le déroulement d'une scène dont le cadre est pré-défini tout en permettant une interaction avec l'utilisateur.

**1.2. Principes artistiques**

L'opéra interactif s'inscrit dans la lignée des œuvres ouvertes proposées par les compositeurs dans les années cinquante et soixante. En 1957, Pierre Boulez (né en 1925) dans sa *Troisième Sonate* pour piano et Karlheinz Stockhausen (né en 1928) dans son *Klavierstück XI*, proposent au pianiste des possibilités d'enchaînements variables entre des sections clairement écrites : l'instrumentiste, en appliquant des règles combinatoires posées par le compositeur, peut proposer des parcours différents dans l'œuvre d'une exécution à l'autre. Mais un autre compositeur, André Boucourechliev (1925-1998) va plus loin que la simple articulation de contenus pré-définis : dans *Archipel IV* (1971) pour piano, il propose au musicien de constituer dynamiquement la musique elle-même, en associant des schémas mélodiques à des schémas rythmiques en temps réel.

Ces œuvres mettent d'une certaine façon l'auditeur hors-jeu. En effet, elles offrent aux interprètes des possibilités de renouvellement de leur rapport à l'œuvre, tandis que le compositeur pressent une nouvelle façon d'écrire, qui insiste plus sur des processus à déployer que sur des contenus à jouer fidèlement. L'auditeur ne se sent pas concerné par ce style d'écriture, qu'il percevra difficilement, sachant qu'il faudrait écouter l'œuvre plusieurs fois pour en percevoir le renouvellement.

Dans l'opéra *Votre Faust* (1968), le compositeur Henri Pousseur et l'écrivain Michel Butor ont tenté de proposer des modalités d'interaction permettant au public d'infléchir le cours de l'histoire. Malheureusement, cette œuvre n'a pu être créée dans des conditions correctes et les moyens d'expression mis à disposition du public (vote, interventions à voix haute, etc.) ont suscité un vaste désordre, qui a définitivement éloigné l'œuvre des scènes lyriques.

Les formes ouvertes « traditionnelles » sont tombées en désuétude. Nous pensons que l'informatique multimédia peut permettre de redonner vie à ce genre, en le transformant au profit de l'auditeur et au détriment du musicien interprète : le premier va pouvoir manipuler des contenus musicaux que le second se contentera d'enregistrer, étant dessaisi de ces possibilités d'intervention. De plus, de notre point de vue, il est important d'affirmer que le CD-ROM peut être le support d'une œuvre musicale à part entière, et pas seulement du commentaire d'une musique comme le sont de nombreux titres multimédia.

### 1.3. Principes interactifs

Les « documents » manipulés par l'utilisateur ont ici une forte spécificité, puisqu'ils sont à caractère artistique. Cela conditionne les modalités de l'interaction. Pour le compositeur-auteur, il s'agit de prendre en compte explicitement l'auditeur/spectateur. Pour ce dernier, il s'agit non pas d'achever une œuvre laissée en suspens, mais bien de participer au déploiement de contenus lyriques, qu'il vient lire et d'une certaine façon écrire ou ré-écrire.

### 1.4. Conséquences sociologiques

Que ce soit pour le compositeur ou pour l'auditeur, il se crée un nouveau rapport à l'opéra. Lecture et écriture ne sont plus séparées comme dans l'opéra traditionnel. Il serait toutefois illusoire de faire croire à l'utilisateur-spectateur qu'il devient « compositeur » par le truchement de cette œuvre. D'une manière générale, les dispositifs numériques véhiculent l'idée selon laquelle chacun peut devenir un artiste, qu'il suffit de mettre en œuvre des procédés ou des « trucs » relevant d'un savoir-faire, que chacun possède les qualités nécessaires potentiellement en lui.

Nous récusons ces affirmations, non point pour protéger la « figure » du compositeur, mais parce que précisément elles ne l'abolissent pas ; bien au contraire elles la manipulent indirectement, car elles promettent à l'utilisateur qu'il pourrait devenir un artiste au sens occidental traditionnel : chacun doit pouvoir atteindre cet « état ». Nous croyons au contraire que l'intervention créative de l'utilisateur dans notre opéra ne fait plus appel à ces conceptions et que le rapport interactif ouvre de nouvelles modalités d'expression qui d'une certaine manière s'affranchissent des classiques divisions entre artistes et spectateurs, entre
« amateurs » et « professionnels ».

## 2. Comparaison entre opéra traditionnel et opéra numérique

**2.1. Approche utilisée**

Nous souhaitons ici étudier et modéliser les interactions qui se produisent lors de la représentation d'un opéra. Une telle approche est fondée sur la notion de système, c'est-à-dire pour nous un ensemble d'intervenants, humains et machines, appelés agents, coopérant dans le cadre d'une activité donnée. Nous avons choisi une méthode de modélisation inspirée de MADEINCOOP [Zacklad 1993, Rousseaux 1995], développée depuis 1992 par un groupe de chercheurs de spécialités différentes travaillant sur le thème de la modélisation de la coopération hommes-machine. Celle-ci s'articule autour de quatre grands principes :

- La coopération est modélisée au niveau des connaissances : des buts et des connaissances sont attribués aux agents humains et artificiels du système étudié, sans faire référence aux caractéristiques techniques.

- La méthode commence par établir un Modèle Global de l'Activité Collective (MOGAC) qui décrit l'organisation, les tâches effectuées et les caractéristiques des agents participants. Ainsi apparaissent les sous-groupes d'agents interagissant fréquemment pour poursuivre un but commun, dont la coopération sera précisée ultérieurement.

- Lors de la modélisation fine de la coopération, la perspective adoptée sera orientée sur les interactions entre systèmes humains et artificiels plutôt que sur leurs procédures et représentations internes.

- Dans la méthode, la modélisation de l'activité de l'acteur se fait selon trois points de vue : le point de vue du modèle de résolution de problème, le point de vue de la coordination, et le point de vue de la communication.

Précisons enfin que le MOGAC que nous venons d'évoquer s'articule en trois parties :

- le modèle de tâche qui donne les buts globaux de l'activité, les moyens, les dépendances chronologiques entre les tâches[2] ;

- le modèle des agents, qui pour chacun d'entre eux, définit ses savoir-faire, ses responsabilités et ses disponibilités ;

- le modèle d'organisation, qui donne l'appariement entre les agents et les tâches, pour définir des situations d'interactions entre agents, auxquels seront adjoints des principes généraux de coordination.

Nous allons appliquer ces principes à la modélisation de la situation de représentation

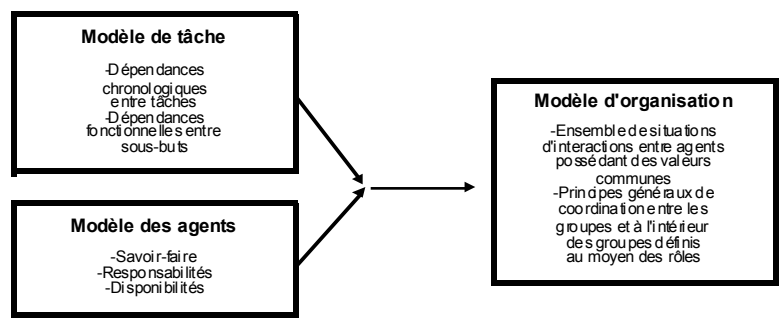

**Figure n° 1.** *Les différents modèles du MOGAC.*

d'opéra.

**2.2. Modélisation de l'opéra traditionnel**

Les agents humains sont répartis entre ceux qui produisent le spectacle (« acteurs » qui « font », et « régisseurs » qui ne « font pas » mais indiquent aux « acteurs » à quel moment faire et suggèrent comment faire) et ceux qui le reçoivent (spectateurs). Les agents artificiels ne sont pas ici liés à la programmation orientée objets, mais sont pour nous des systèmes capables de jouer un rôle de collaborateur intelligent. Il s'agit dans le cas de l'opéra traditionnel des systèmes de scène permettant la gestion des machineries et des éclairages.

---

[2] Dans la méthode MADEINCOOP, le modèle de tâche comporte également une description des dépendances fonctionnelles entre les sous-buts, que nous ne reprenons pas ici, car elle est peu opérante dans le contexte que nous étudions.

Les buts de l'activité opératique relèvent de la célébration d'un rituel dont les habitués témoignent qu'il leur a parfois ouvert les portes de moments rares et intenses [Bonardi 2000]. Selon Michel Poizat [Poizat 1986], il s'agit de produire du chant pour aboutir à une « jouissance lyrique » qu'il rapporte comme Claude Lévi-Strauss à la douleur de la perte du langage. Ainsi, la voix est conduite en permanence du chant dont le contenu sémantique est parfaitement intelligible jusqu'au son proche du cri qui masque le contenu verbal. Produire du chant ne peut se réduire à chanter, à placer des syllabes sur des hauteurs ; il s'agit, bien au-delà, de piloter plusieurs « couples de variables » rendant compte de dialectiques : intelligibilité verbale/son musical, couleur claire (énergie déplacée sur les harmoniques plus élevés)/couleur sombre (énergie recentrée sur la fondamentale), de construire des continuités/discontinuités dans l'émission vocale, etc.

Dans l'opéra, les tâches sont divisées en deux sortes : celle qui concerne les agents « artistes », à savoir l'interprétation d'une partition au sens large (partition musicale, feuille de route d'éclairages, etc.) ; d'autre part, celles qui concernent tous les agents, « acteurs », « régisseurs », « spectateurs », consistant à observer la représentation, à interpréter ce qui est vu et entendu, et enfin à actualiser certaines conceptions a priori. Ces dernières peuvent être aussi bien un point de mise en scène, la mise en place d'un passage musical que l'idée qu'un spectateur se fait a priori d'un opéra[3], ou encore la représentation qu'a un chanteur de sa propre voix.

Le tableau de la figure n° 2 rend compte de ces différents aspects.

---

[3] à l'instar du bouleversement mis en scène par Agnès Jaoui dans son film *Le goût des autres* (2000), qui voit la vie du personnage principal basculer à l'occasion d'une représentation de *Bérénice* de Racine.

| | |
|---|---|
| **Agents humains** | <ul><li>des « acteurs » produisant des contenus visuels ou sonores (chanteurs, choristes, machinistes, éclairagistes, etc.),</li><li>des « régisseurs » prescrivant et synchronisant l'activité des artistes (chef d'orchestre, metteur en scène et ses assistants, régisseur plateau, etc.),</li><li>des spectateurs.</li></ul> |
| **Agents artificiels** | systèmes de gestion des machineries et des éclairages (« système de scène »). |
| **Buts** | <ul><li>célébrer un rituel,</li><li>produire du « chant »,</li><li>vivre intensément.</li></ul> |
| **Moyens pour atteindre ces buts** | <ul><li>des espaces pour les artistes (scène, fosse d'orchestre, loges, coulisses, cabines de supervision, etc.),</li><li>des espaces pour les spectateurs (salle, foyer, hall d'entrée, etc.),</li><li>des corps chantants (qui se confondent avec certains agents « acteurs »),</li><li>des instruments,</li><li>des partitions et des indications (écrites et orales)</li><li>des machines de scène,</li><li>des dispositifs de contrôle (caméra sur scène montrant le chef d'orchestre, jumelles des spectateurs, etc.)</li></ul> |
| **Tâches** | <ul><li>exécuter une indication,</li><li>interpréter une partition,</li><li>suivre la représentation et son évolution,</li><li>interpréter ce qui est vu et entendu,</li><li>prescrire le jeu des agents « acteurs » et le comportement des agents artificiels,</li><li>actualiser des conceptions a priori (de l'œuvre mais aussi de soi-même).</li></ul> |
| **Dépendances chronologiques entre tâches** | synchronisation. |

**Figure n° 2.** *Modèle global de la tâche de la représentation d'opéra.*

Tentons maintenant d'associer les tâches aux agents et vice-versa. Pour améliorer la lisibilité des schémas, nous avons associé à chaque type d'agent une image : un visage de chanteuse pour les « acteurs », une main manipulant une baguette de direction pour les « régisseurs », une photographie d'un véritable système de scène pour les agents artificiels, et un visage de spectatrice reproduit plusieurs fois pour les « spectateurs ». Voici le résultat obtenu en partant des agents :

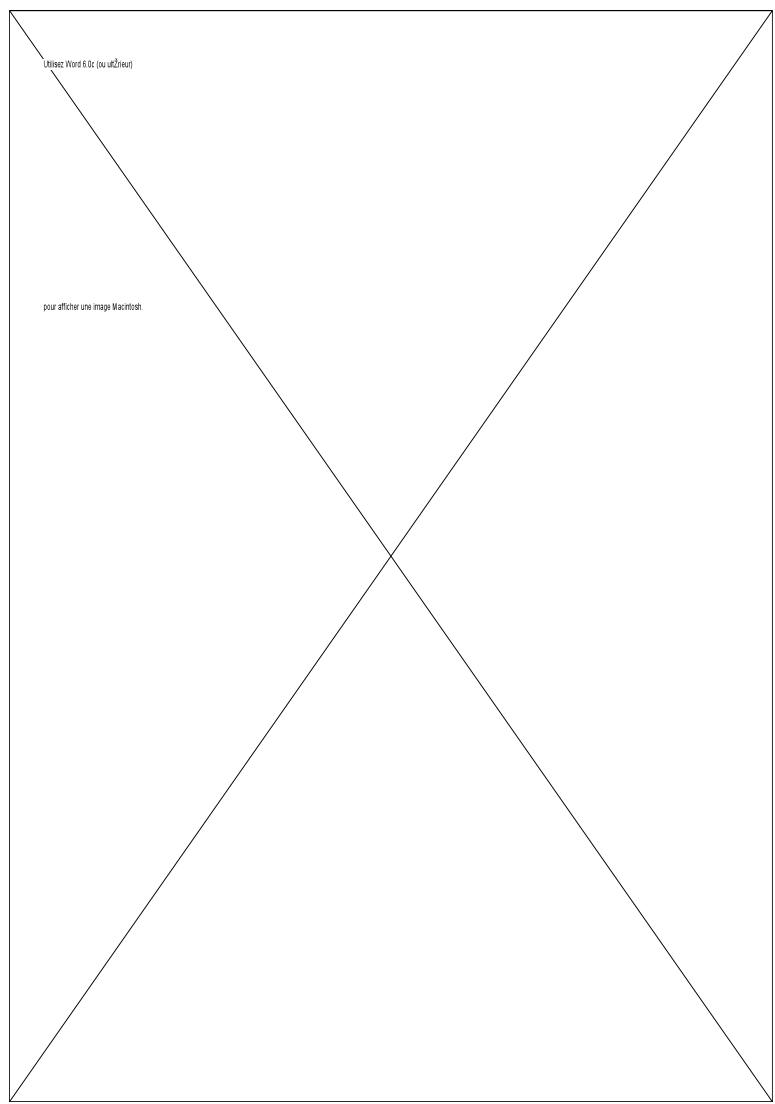

**Figure n° 3.** *Tableau des tâches confiées aux différents agents.*

Le tableau précédent montre que les tâches peuvent être réparties selon quatre types : tâches de fond, tâches de prescription, tâches d'interprétation et tâches de mise à jour.

La situation de représentation de l'opéra peut être modélisée finement en considérant successivement la communication entre agents, leur coordination tout en rendant compte de la résolution collective de problèmes, qui dans le cas d'un opéra est fondée sur les tâches d'interprétation des partitions et des situations. La figure n° 4 représente nos modèle de résolution collective de problème et de communication dans l'opéra[4] :

---

[4] Les autres modèles peuvent être consultés dans notre thèse : BONARDI, Alain, *Contribution à l'établissement d'un genre : l'Opéra Virtuel Interactif*, thèse de l'Université Paris IV, 2000 (plus particulièrement le chapitre 4).

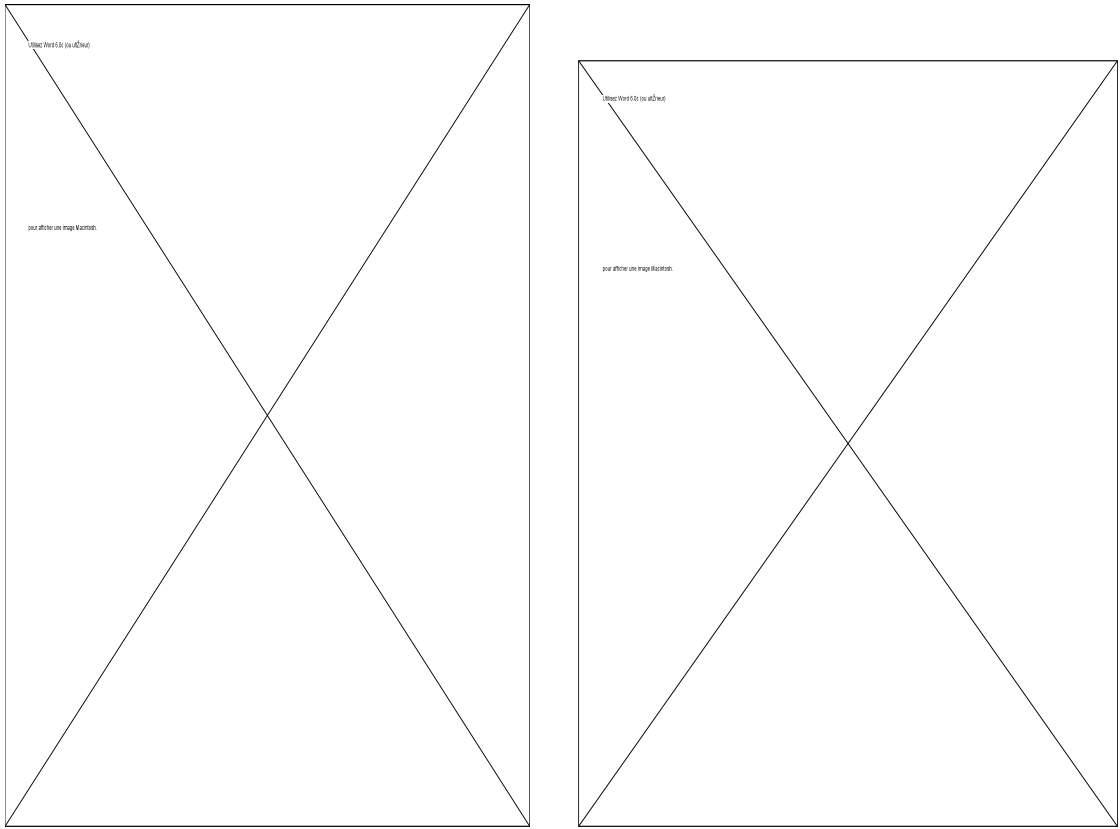

**Figure n° 4.** *Modèles de résolution collective de problème (à gauche)
et de communication dans la représentation d'opéra (à droite).*

Le modèle de communication traduit la nécessité d'un ou plusieurs langages de communication supportés par des médias adéquats. La communication est ici multimodale (son émis par les acteurs en direction des autres agents à l'exception des « systèmes de scène », image des acteurs, contact tactile avec les systèmes de commande).

### 2.3. Modélisation de l'opéra interactif

Nous nous posons désormais la question de la modélisation de la coopération entre agents lors de la « représentation » d'opéra interactif. Nous emploierons dans la suite de notre raisonnement l'acronyme OI pour désigner un opéra interactif.. La représentation de l'OI donne lieu au face à face d'un ordinateur et d'un spectateur/utilisateur. Par rapport à un opéra traditionnel, le spectateur participe à un double processus : il se situe à la fois à l'extérieur de l'œuvre, dont l'illusion de construction l'éloigne, et à l'intérieur en participant à sa constitution.

La première question que nous pouvons poser concerne les quatre types d'agents de l'opéra traditionnel : que deviennent-ils dans l'OI ? Pour y répondre, nous devons tout d'abord mieux caractériser les deux situations d'interaction. Alors que le spectateur d'opéra est absent de ce qui se joue sur scène, l'utilisateur d'un ordinateur se trouve d'emblée en situation de dialogue avec sa machine. La traditionnelle représentation d'opéra est une expérience de diversité à laquelle s'expose le spectateur (c'est l'expérience « qu'il existe du divers ») ; ce divers lui est parfois accessible et pas si divers que cela. En revanche, l'interaction avec un ordinateur est une expérience d'altérité, qui passe par l'attribution de sens à des dispositifs tels qu'un prompteur ou des icônes, dans le but d'investir d'esprit les processus qui les manipulent. C'est donc sous le signe du dialogue créatif que vont se distribuer les rôles dans l'OI.

| AGENTS DE L'OPERA TRADITIONNEL | AGENT TENANT CE ROLE DANS L'OI | PRISE EN CHARGE DU ROLE |
|---|---|---|
| « Acteurs » (agents humains) | Ordinateur (agent artificiel) / Spectateur (agent humain) | L'ordinateur produit des contenus visuels et sonores, et veut suggérer qu'ils proviennent d'acteurs « artificiels » auxquels le spectateur les associe (avatars, personnages, icônes, etc.)<br><br>L'utilisateur joue « à l'opéra » et peut produire des contenus visuels et sonores. |
| « Régisseurs » (agents humains) | Ordinateur (agent artificiel) / Spectateur (agent humain) | L'ordinateur exécute des séquences d'instruction coordonnant les actions des différents « acteurs » multimédia.<br><br>Par les mécanismes de forme ouverte mis à disposition du spectateur, ce dernier assure un rôle de régisseur. |
| « Spectateurs » (agents humains) | Spectateur (agent humain) | Le spectateur interagit avec le contenu multimédia. |
| « Système de scène » (agents artificiels) | Ordinateur (agent artificiel) | L'ordinateur multimédia est tout entier système de scène. |

**Figure n° 5.** *Correspondances de rôles entre les typologies d'agents (opéra traditionnel et OI).*

Ainsi, les « acteurs » humains disparaissent au profit de représentations comme des avatars ou des personnages animés, auxquels l'utilisateur attribuera l'origine des contenus visuels et sonores joués par l'ordinateur. Le spectateur est aussi amené à jouer un rôle d'« acteur », puisqu'il peut par son interaction provoquer l'émission de sons et d'images. Le rôle de « régisseur » est également partagé entre l'ordinateur et le spectateur : d'un côté, la machine contrôle ses différents organes multimédia; de l'autre, le spectateur, à qui il est proposé de participer à une sorte de « forme musicale ouverte » sur ordinateur, joue aussi le rôle de « régisseur ». De plus, l'agent « spectateur » demeure, mais, comme nous venons de le dire, il interagit désormais avec le dispositif scénique. Enfin, les « systèmes de scène » disparaissent en tant que tels pour se confondre avec l'ordinateur lui-même[5]. Nous avons repris l'ensemble de ces remarques dans le tableau de la figure 5.

Nous ne pouvons nous contenter de projeter ce que nous connaissons de l'opéra traditionnel sur l'OI. Réciproquement, nous devons nous demander s'il n'existe pas des rôles qui ne figureraient pas dans l'opéra traditionnel mais apparaîtraient dans l'OI. C'est le cas de l'activité de composition. Cette dernière n'est plus séparable de la réalisation de l'œuvre : une partie s'écrit au moment de l'exécution, et d'une certaine façon le spectateur/utilisateur contribue à la partition que joue l'ordinateur.

Nous pouvons ainsi établir un modèle global de tâche de la représentation d'OI, tel que présenté à la figure 6.

---

[5] La scène est alors l'écran de l'ordinateur, qui impose ses contraintes de taille, d'affichage et de résolution.

| Agent humain | un utilisateur / spectateur jouant avec un opéra. |
|---|---|
| Agent artificiel | un ordinateur support de l'interaction créative. |
| Buts pour l'agent humain | <ul><li>jouer avec l'opéra,</li><li>constituer de l'altérité,</li><li>s'exposer à une diversité que l'on contrôle et/ou génère.</li></ul> |
| Buts pour l'agent artificiel | <ul><li>« donner le change » au spectateur, en offrant une interaction à la fois plausible et intéressante,</li><li>mettre en œuvre la représentation d'OI,</li></ul> |
| Moyens pour atteindre ces buts | <ul><li>des moyens de production visuelle et sonore,</li><li>des séquences d'instructions (animation multimédia),</li><li>des programmes modélisant l'interaction,</li><li>des dispositifs de contrôle (souris, clavier, etc.).</li></ul> |
| Tâches | <ul><li>exécuter une instruction,</li><li>suivre la représentation et son évolution,</li><li>prescrire l'évolution de la représentation,</li><li>prescrire des affichages et des diffusions sonores,</li><li>interpréter ce qui est vu et entendu,</li><li>interpréter l'action du spectateur,</li><li>actualiser des conceptions a priori (de l'œuvre mais aussi de soi-même).</li></ul> |
| Dépendances chronologiques entre tâches | synchronisation. |

**Figure n° 6.** *Modèle global de la tâche de la représentation d'OI.*

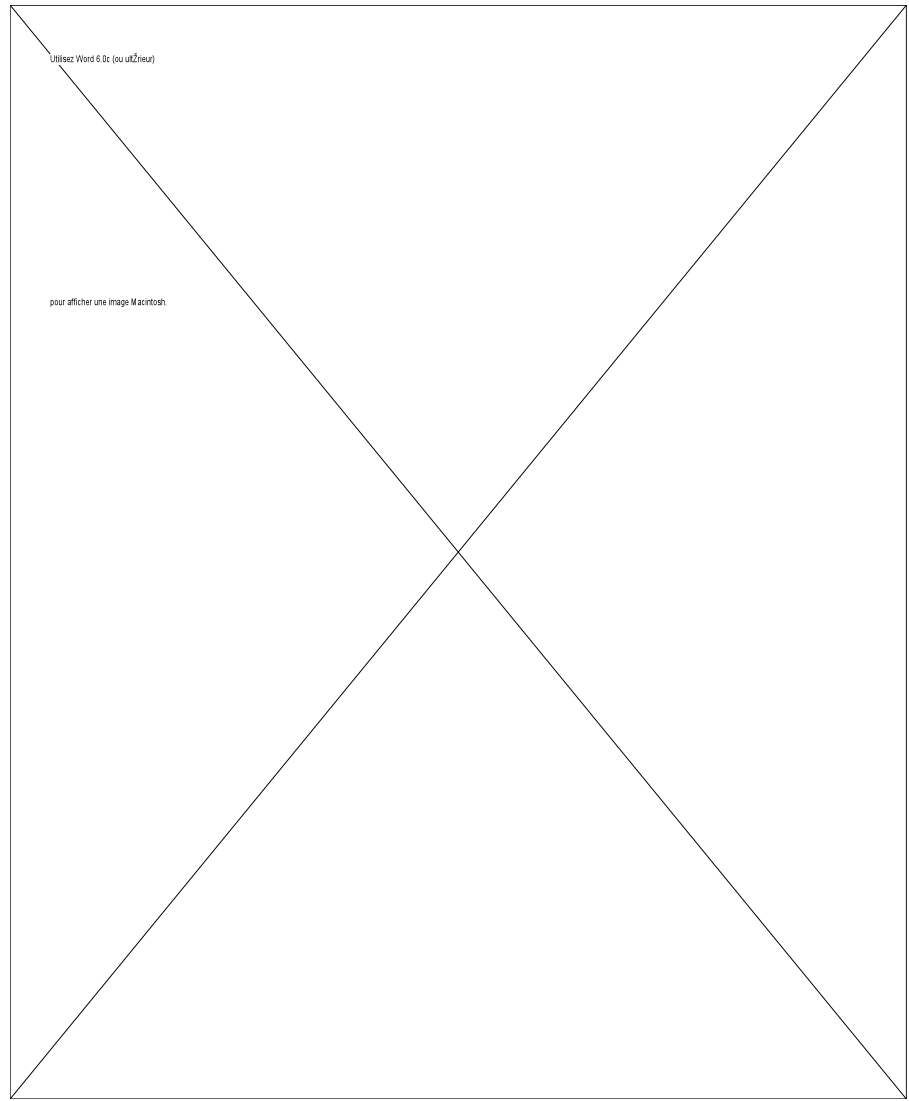

**Figure n° 7.** *Modèle d'appariement des agents et des tâches dans l'OI en partant des agents.*

## 3. Modèles de l'interactivité dans *Virtualis*

### 3.1. Principes généraux

Nous présentons maintenant les principes d'interactivité que nous avons retenus pour déployer les modèles que nous venons de présenter et permettre à l'utilisateur de « jouer les rôles » que nous avons décrits. Quelles sont les grandes orientations que nous avons choisies ?

- nous cherchons une interactivité qui ne se dirait pas comme telle. Les liens et icônes se déploient habituellement pour désigner l'interactivité et ses lieux. Nous souhaitons mettre au point des processus fluides où il est implicitement proposé à l'utilisateur d'intervenir, s'il le souhaite. S'il ne le fait pas, l'œuvre continue sa trajectoire, selon des données de base et selon ce que l'ordinateur a mémorisé des parcours antérieurs. Il s'agit ainsi de ne pas interrompre le flux de l'œuvre, mais plutôt de l'orienter.

- comme dans toute application interactive, il se pose la question de la prise en compte de l'intention de l'utilisateur et de son couplage à la réponse de la machine. Dans *Virtualis*, nous avons essayé de mettre en œuvre un modèle non psychologique : les motivations et le comportement de l'utilisateur ne sont pas modélisés. Ce dernier est considéré comme un élément extérieur qui peut agir sur le système autonome que constitue l'œuvre interactive sans être explicitement pris en compte dans sa complexité psychologique. Pour ce faire, nous avons élaboré un modèle fondé sur des forces physiques.

L'opéra interactif *Virtualis* propose trois types de scène :

- des tableaux offrant des interactions ludiques sur des dialectiques de l'opéra. Ainsi, dans le tableau intitulé « Les mots et la mer », des rochers représentent des mots alors que la mer représente la musique, et selon le niveau d'eau, ajusté par l'utilisateur, les mots parlés sont plus ou moins altérés, leur contenu sonore étant progressivement transformé en contenu musical. Nous ne reviendrons pas plus avant sur ces tableaux dans cet article.

- des scènes de transition intitulées « parcours de la musique » qui donnent l'occasion au joueur d'évoluer dans un univers tri-dimensionnel où la musique est représentée sous forme de métaphores graphiques, permettant à l'utilisateur d'interagir avec des contenus musicaux associés à des objets tri-dimensionnels. Nous présenterons ces scènes au paragraphe 3.2.

- des scène de transition intitulées « le Récit », qui sont des courts moments narratifs interactifs entre les personnages. Ce sont les passages pour lesquels nous avons développé le modèle physique sur lequel nous reviendrons en détail au paragraphe 3.3.

**3.2. Les « parcours de la musique »**

Les « parcours de la musique » sont des transitions entre tableaux. Elles peuvent intervenir ou non, selon un choix aléatoire, entre ces derniers. Leur principe demeure toutefois identique à chaque fois, bien que les musiques qui y prennent part varient. Trois fonctions principales sont proposées à l'utilisateur :

- « Errer » à travers la musique, en se promenant dans un espace géométrique qui en représente certaines propriétés, dans un univers où désormais musique et paysage se confondent.

- Modifier la musique, représentée sous forme d'objets graphiques qu'il peut manipuler.

- S'orienter vers telle ou telle autre séquence musicale.

Examinons tout d'abord les principes de la « promenade » dans les contenus musicaux. L'utilisateur, en quelque sorte fixé à la caméra qui nous permet de voir le paysage, « vole » comme un oiseau dans une vaste cage cubique, dont l'idée provient de l'univers du surprenant film *Cube*[6], réalisé par Vicenzo Natali (1999). Cet espace est occupé par des objets musicaux représentés par des volumes tri-dimensionnels plus ou moins cachés, aux formes variées. Ces objets sont des fragments musicaux monodiques ; ils peuvent par exemple provenir de l'éclatement d'un morceau polyphonique en lignes individuelles. La figure 8 montre une vue de dessus du dispositif.

---

[6] Dans ce film, un groupe de personnages se réveille dans un univers constitué de salles cubiques qui recèlent des pièges. Ils sont en quelque sorte prisonniers de la géométrie.

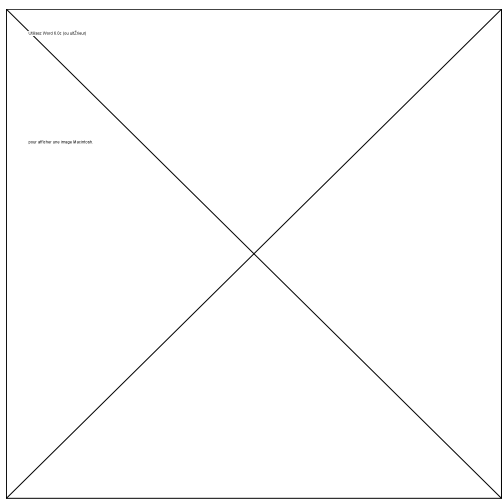

**Figure n° 8.** *Vue de dessus (en plongée) du dispositif des « parcours dans la musique ». L'utilisateur peut descendre et s'approcher d'un des quatre fragments monodiques, qu'il peut parcourir en détail.*

Les musiques associées à ces objets sont jouées en boucle, chacune ayant sa propre longueur, ce qui conduit à un décalage permanent entre elles. Les fichiers sonores se mélangent, mais les intensités relatives dépendent de la distance et de la position de l'utilisateur par rapport aux entités graphiques qui les représentent. Lorsqu'il se trouve à l'intérieur d'un volume associé à un objet musical, l'utilisateur peut décider de s'intéresser plus spécifiquement à lui. Par une commande, il choisit de se laisser porter par la musique, c'est-à-dire qu'il n'est plus libre de ses mouvements, mais est entraîné par le flot de la musique à travers l'objet qui la représente. La figure 9 ci-dessous montre un exemple de parcours dans un objet tri-dimensionnel ayant la forme d'un tunnel, associé à une mélodie jouée au piano.

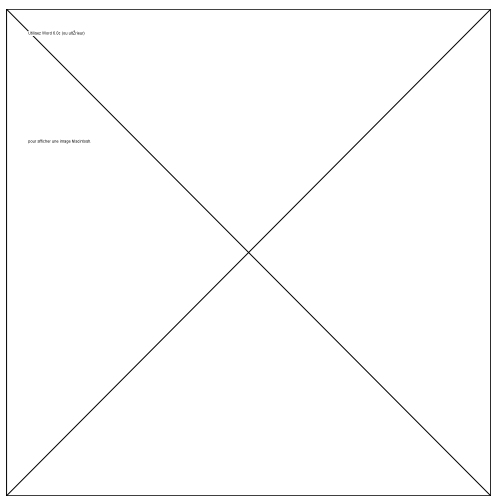

**Figure n° 9**. *Déplacement au rythme de la musique à l'intérieur d'une représentation d'un fragment monodique.*

Pour confectionner ces parcours, nous avons tout d'abord créé une application nommée ALMA, qui permet d'associer des objets graphiques à des contenus musicaux, de créer des variations mélodiques sur un objet donné et d'implémenter des interactions. Dans la version MIDI, l'utilisateur peut non seulement modifier des paramètres macroscopiques comme le tempo, le volume de restitution sonore, ou la position (panoramique) du son sur l'échelle stéréo de gauche à droite, mais aussi générer des variations structurales qui modifient les contenus joués. L'utilisateur choisit un fragment monodique associé à un objet graphique, et demande à l'ordinateur d'en générer des variations. Le système utilise le principe de recouvrement d'une entité mélodique par les motifs qui la composent [Baboni-Schilingi 1998].

## 3.3. Le « Récit »

Le « Récit » met en scène deux personnages, un homme et une femme, racontant une histoire donnée articulée en très courts « moments ». La question est de savoir comment rendre ce « Récit » interactif, et donc de choisir un modèle d'interaction pertinent. Nous avons choisi un modèle physique, utilisant des forces équivalent à la force d'attraction électrique et les champs de forces associés, en reliant, involontairement ou non, conjonctions amoureuses et conjonctions particulières voire conjonctions planétaires[7], l'histoire ne disant pas si la pomme d'Isaac Newton avait le même goût que celle que croquèrent Adam et Eve.

Ce récitatif se déroule à l'écran dans un espace clos, où évoluent les deux personnages. Notre idée est celle d'un duo plus « chorégraphique » que « théâtral ». En effet, le point de vue de la danse, où le mouvement crée l'expression s'accorde bien au modèle physique retenu qui s'intéresse aux positions et vitesses. Nous préférons utiliser des poses sur des mouvements précis comme celles de la figure 10, qui seront soit utilisées telles quelles, soit montées avec une certaine vitesse pour donner de courtes animations qui alterneront avec des phases statiques. Il s'agit d'intégrer ces aspects visuels au modèle de forces qui régit le « Récit ».

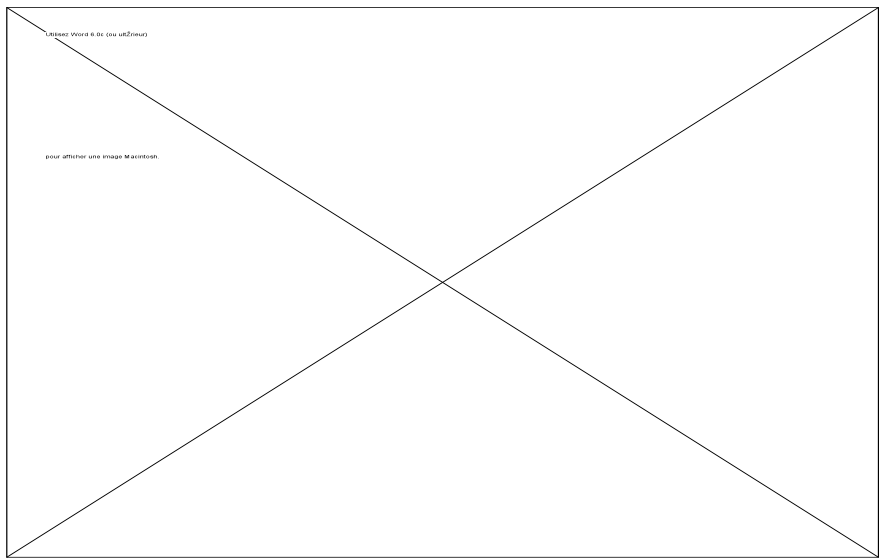

**Figure n° 10.** *Exemples de positions de l'un des deux modèles.*

Les forces physiques sont exercées d'une part par chacun des deux personnages, d'autre part par des attracteurs situés en dehors de l'écran, produisant des champs de force supposés constants s'exerçant sur les personnages[8]. L'utilisateur/spectateur dispose des possibilités suivantes :

- Il peut d'une certaine façon diriger les chanteurs, en choisissant de donner un départ à l'un des deux chanteurs, en cliquant sur son icône. Mais il ne suffit pas de cliquer une fois, il faut régulièrement relancer le personnage dans son chant et son jeu, faute de quoi ce dernier s'estompe peu à peu visuellement et sur le plan sonore.

---

[7] Rappelons que les forces électrique entre deux particules chargées et la force de gravitation entre deux planètes ont des expressions analogues en ☒, r étant la distance entre les deux entités en présence.

[8] Sur le détail du modèle d'interaction, nous renvoyons le lecteur au paragraphe suivant.

- Il peut déplacer l'un des deux chanteurs, ce qui peut provoquer la modification de ce que celui-ci chante et le déplacement de l'autre chanteur.

- Il peut choisir un élément de décor menant à un autre court « moment » du « Récit » (sinon, par défaut, l'ordinateur se rendra au « moment » suivant) selon un principe de forces physiques qui sera explicité au paragraphe suivant.

- Il peut bifurquer vers un autre tableau.

Toute la dynamique du tableau est fondée sur le calcul en temps réel des forces et la mise à jour de la position et du chant des deux personnages. Nous supposons pour ce faire que quatre forces du type des forces électriques ou de gravitation s'exercent. Chaque personnage est donc modélisé par quatre « charges » ou « masses » que nous pourrions qualifier d' « affectives », placées selon quatre axes : aspiration à la tendresse, audace / résignation, égoïsme et jalousie. Elles expriment soit une grandeur positive, équivalent d'une masse, soit une grandeur (positive) et une nature (positive/négative), ce qui équivaut à une charge. Ces axes sont invariants, mais le poids de chacun des personnages sur chacun des axes change selon le moment du « Récit ».

Les interactions possibles concernent :
- l'attraction / répulsion entre ces deux personnages ;

- l'attraction / répulsion entre chacun des personnages et les forces extérieures.

La figure 11 ci-dessous montre les différents champs de forces et forces qui s'exercent sur la scène représentée par un rectangle. Les forces $F_{21}$ et $F_{12}$ sont respectivement celle qu'exerce l'homme sur la femme et son opposé. $G_1$ et $G_2$ quant à eux représentent des champs de forces s'exerçant sur les deux personnages selon les quatre « masses » ou « charges » que nous avons évoquées.

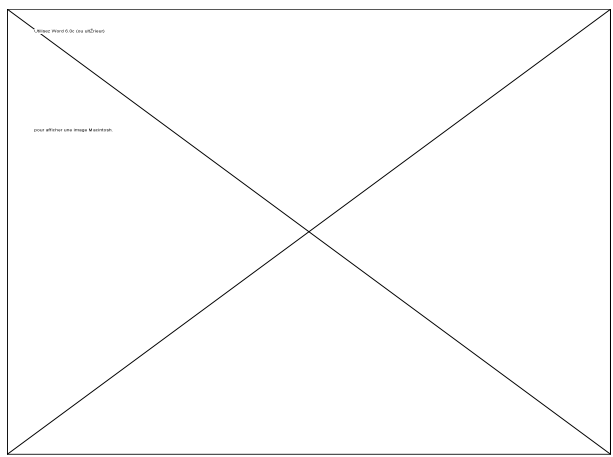

**Figure n° 11.** *Forces internes et externes s'exerçant sur les personnages.*

Pour ajuster les positions des deux personnages, l'ordinateur résout pour chacun d'eux l'équation fondamentale de la dynamique, c'est-à-dire pour le personnage féminin : 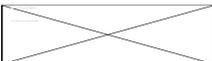 et pour le personnage masculin : 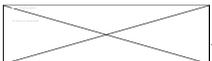. L'algorithme d'Euler est utilisé deux fois pour une double intégration qui permet de calculer les coordonnées des deux modèles.

Le modèle de forces régit également la variation du texte et de la musique. D'une certaine façon, comme en danse, le mouvement crée l'expression. Chaque séquence chantée a donné lieu à plusieurs variantes à la fois du point de vue du texte et du point de vue de la musique. Les variantes textuelles ont été obtenues par glissement sémantique progressif sur un thème, en allant soit des sentiments intérieurs du personnage vers l'extérieur, soit dans l'autre sens. Sont ainsi constitués des axes sémantiques $AS_i$ sur lesquels les forces extérieures au personnage auquel l'axe est associé ont plus ou moins de prise, selon un coefficient 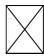 compris entre 0 et 1[9].

Donnons l'exemple d'un moment dramatique qui se passe à la campagne. La thématique se résume en quelques phrases :

L'homme et la femme passent l'après-midi à la campagne.

---

[9] En fait, ces coefficients sont des cosinus d'angles compris entre 0 et 90 degrés, indiquant la direction de l'axe sémantique concerné, et permettant donc de calculer le produit scalaire des vecteurs unitaires de chaque axe avec chacune des forces.

L'homme aime ce lieu ; la femme s'ennuie. L'homme veut rester. La femme ne sait pas ce qu'elle veut. Elle soupçonne de ne pas être la première à venir avec l'homme en ce lieu.

L'écran présente une gradation de gauche à droite : du plus campagnard à ce qui est le plus urbain, le tout rassemblé sur le même ruban qui peut défiler, selon l'endroit où se situent les personnages.

La figure 12 présente deux exemples d'axes sémantiques empruntés par les personnages.

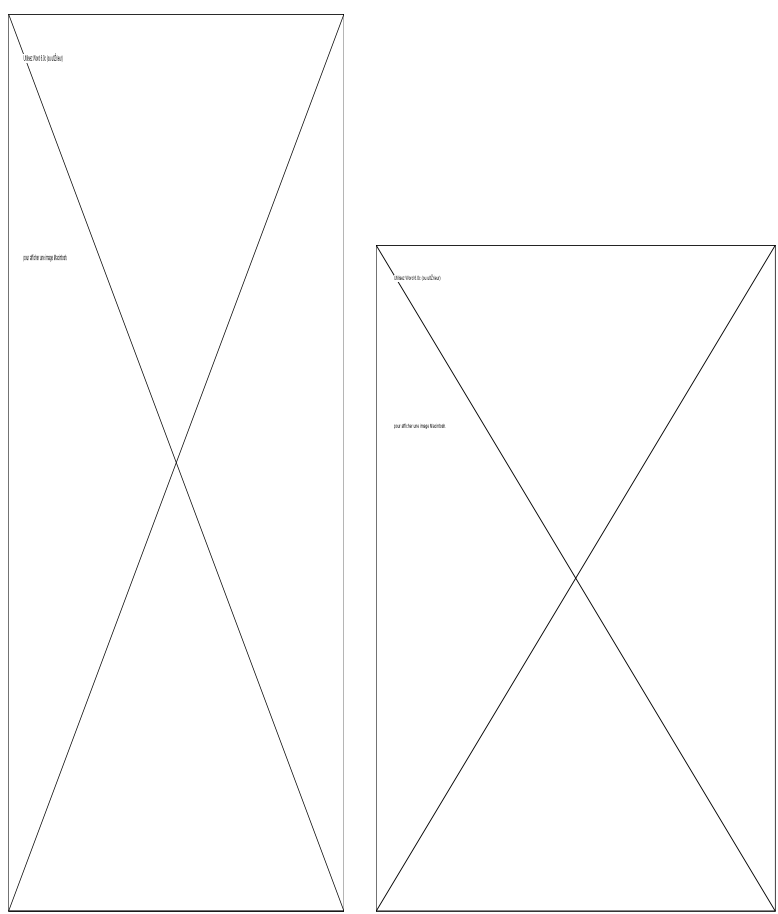

**Figure n° 12.** *Deux axes sémantiques (à gauche, pour la femme ; à droite pour l'homme).*

Par le même procédé sont constitués des axes musicaux, $AM_j$, de la mélodie initiale à sa variation la plus éloignée, également sensibles aux champs de forces extérieurs selon un coefficient ⊠. Ces axes sont orthogonaux aux précédents, donc indépendants des niveaux sémantiques. L'ensemble articulé selon deux dimensions constitue donc un réseau musico-textuel, dont nous donnons un exemple ci-dessous, avec l'axe sémantique « elle me plaît aujourd'hui » croisé des variantes musicales. Pour chaque phrase (par exemple « tu as mis le haut que j'aime ») ont été composées (à « la main » ou avec l'aide de l'ordinateur, en utilisant l'environnement OpenMusic, logiciel du forum Ircam) plusieurs variations regroupées selon un axe vertical et classées de la moins à la plus passionnée.

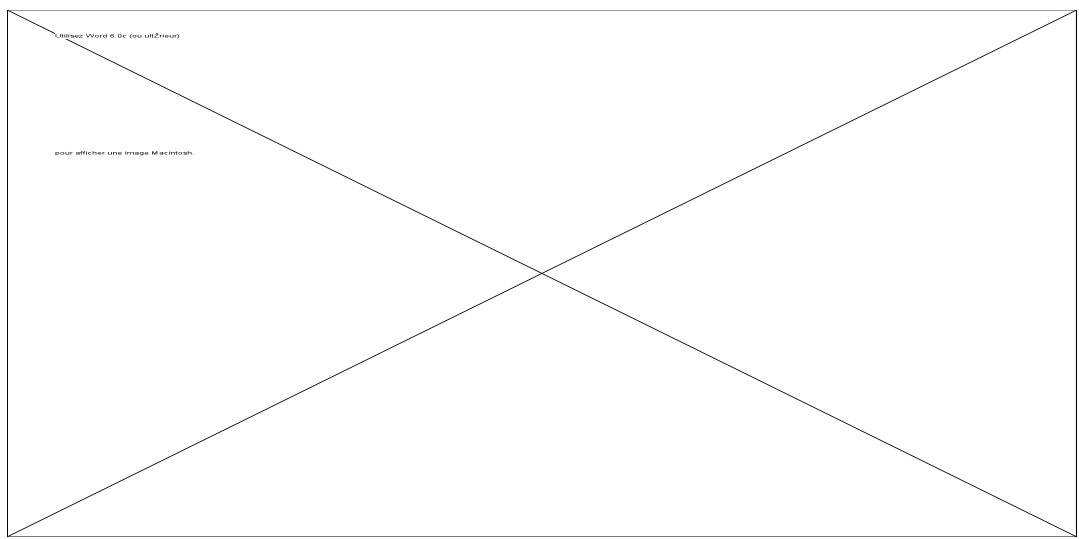

**Figure n° 13.** *Exemple de réseau musico-textuel.*

Voici par exemple (figure 14) trois réalisations musicales de la même phrase confiée au soprano (« moi, je vais tout lui raconter ») :

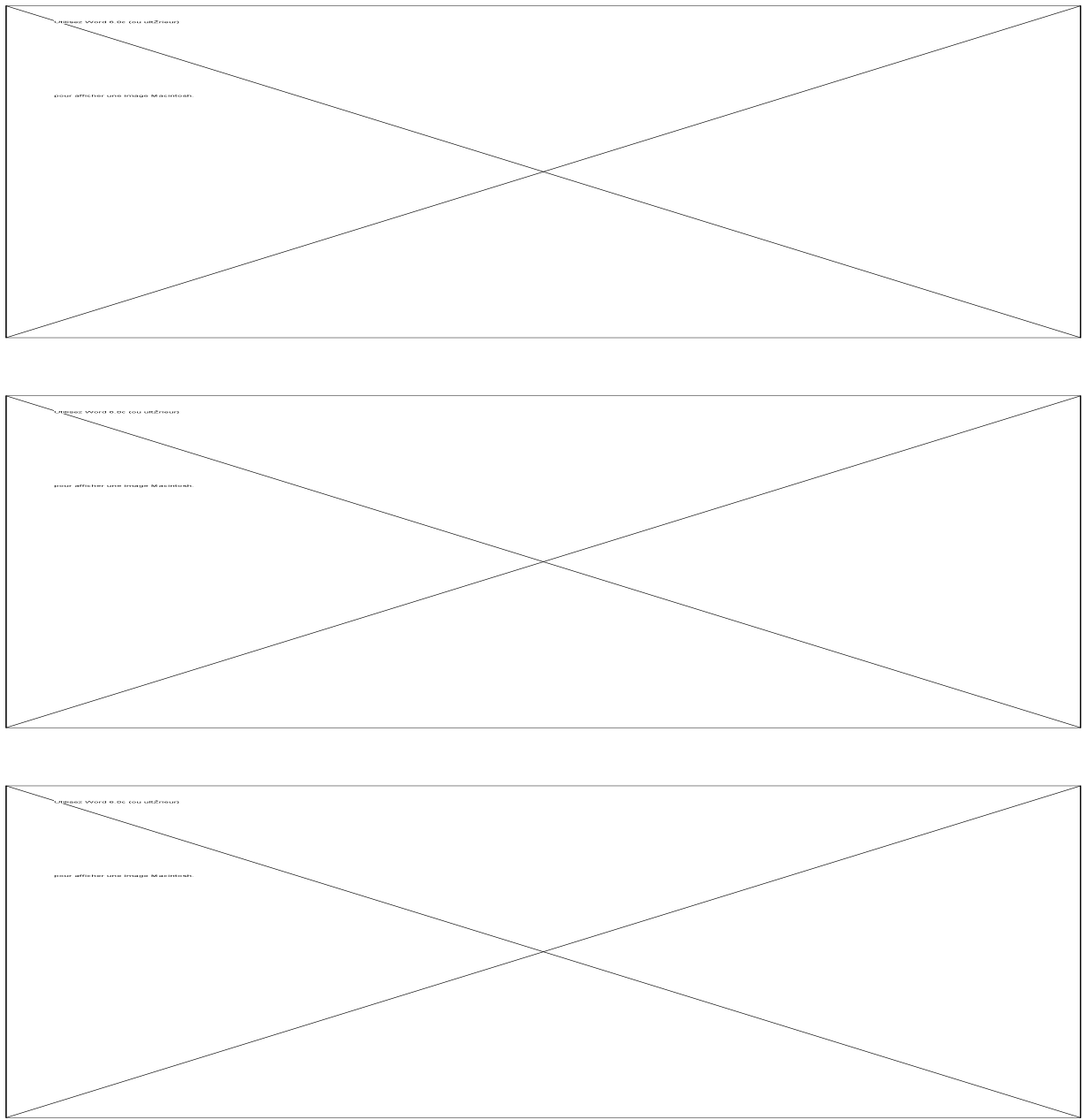

**Figure n°14.** *Trois exemples de réalisation d'une phrase confiée au soprano.*

L'orientation de ce réseau, et notamment le choix d'une progression plutôt selon le texte ou la musique est donné par une variable globale ⊠, sur laquelle l'utilisateur peut agir à tout moment, comme le montre la figure 15.

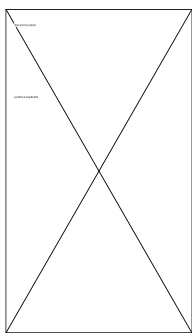

**Figure n° 15.** *Curseur de réglage des influences des champs de forces extérieures entre texte et musique.*

Cette variable varie entre 0 (tout texte) et 1 (tout musique). A tout moment l'axe musical $AM_j$ où se trouve l'utilisateur dans le réseau, déjà doté du coefficient ☒ est pondéré par ☒ et l'axe sémantique $AS_i$, déjà affecté du coefficient ☒, par ☒.

Du point de vue de la réalisation, les différents fragments musicaux du réseau ont été enregistrées par une soprano et un ténor[10]. L'accompagnement des chanteurs a été généré par ordinateur grâce au logiciel de synthèse par modèle physique Modalys (logiciel du forum Ircam). Il est fondé sur des nappes sonores conçues pour fonctionner en fondu-enchaîné lors des changements de climat imposés par les changements de position des personnages et donc des changements de phrases chantées.

Ces modèles physiques mettent directement en relation les entrées gestuelles provoquées par l'utilisateur (mouvements de la souris, etc.) et l'actualisation des contenus. Contrairement aux modèles d'inspiration psychologique, ils abandonnent l'ambition de constituer un niveau supérieur d'interprétation des phénomènes physiques détectés. Ils ne tentent pas d'établir des faits déduits de ces actions, puis de décider de la réaction à apporter à ces faits « calculés », ce qui pose inévitablement le problème de l'interprétation conduisant à devoir expliciter le modèle de la machine pour qu'il adapte son comportement à elle.

### 3.4. Dynamique des enchaînements entre scènes

Nous avons précisé les spécifications des scènes individuelles, mais nous n'avons pas encore examiné en détail comment se produisent les enchaînements entre elles.

[A COMPLETER]

## Conclusion

Nous avons présenté le projet d'opéra interactif *Virtualis* et les modèles physiques d'interaction que nous avons été amenés à développer dans ce cadre. Pour l'utilisateur, ils constituent une activité originale de lecture, d'appropriation et d'écriture de « documents » lyriques, où l'absence de modèle psychologique suscite une relation singulière à une machine dont il ne s'agirait plus de deviner comment elle prend en compte l'utilisateur. Pour le créateur multimédia, les modèles de forces conduisent à une conception informatique qui n'est plus procédurale, mais, à l'instar de la programmation par contraintes, est fondée sur la définition de cadres interactifs au sein desquels l'ordinateur propose des solutions renouvelées d'une utilisation à l'autre.

## Références

---

[10] Nous voudrions remercier ici Sylvie Robert et Eric Gourouben.